\documentclass[prd,12pt,nofootinbib,superscriptaddress]{revtex4-2}
\usepackage{bm}
\usepackage{amsmath}
\usepackage{amssymb}
\usepackage{graphicx}
\usepackage{subfigure}
\usepackage{hyperref}
\usepackage{color}
\usepackage{CJK}
\usepackage{orcidlink}
\usepackage{array}
\usepackage{indentfirst}

\hypersetup{
	colorlinks=true,
	linkcolor=red,
	citecolor=blue,
}

\newcommand{\delay}[1]{\mathbf{D}_{#1}}

\begin{document}

\begin{CJK*}{UTF8}{gbsn}
\title{The constraint on modified black holes with extreme mass ratio inspirals} 	

\author{Chao Zhang  \orcidlink{0000-0001-8829-1591}}
\email{zhangchao666@sjtu.edu.cn}
\affiliation{Department of Physics, School of Physical Science and Technology, Ningbo University, Ningbo, Zhejiang 315211, China}

\author{Guoyang Fu  \orcidlink{0000-0002-7944-8611}}
\email{fuguoyangedu@sjtu.edu.cn}
\affiliation{School of Physics and Astronomy, Shanghai Jiao Tong University, 800 Dongchuan Rd, Shanghai 200240, China}

\author{Yungui Gong  \orcidlink{0000-0001-5065-2259}}
\email{Corresponding author. gongyungui@nbu.edu.cn}
\affiliation{Department of Physics, School of Physical Science and Technology, Ningbo University, Ningbo, Zhejiang 315211, China}

\begin{abstract}
The low-energy effective action of String Theory introduces corrections to the dilaton-graviton sector, resulting in deformed black holes beyond general relativity.
We analyze extreme mass-ratio inspiral systems (EMRIs), where a stellar-mass object spirals into a slowly rotating supermassive black hole including a distinct deviation parameter.
This study examines the effects of this deformation on gravitational wave fluxes, orbital evolution, and phase dynamics, incorporating leading-order post-Newtonian corrections.
With one-year observations of EMRIs, we employ the Fisher information matrix method to evaluate the potential for detecting deviations from general relativity through space-based gravitational wave detectors that utilize time-delay interferometry to suppress laser noise.
The constraint on modified black holes, $\Delta\alpha \preceq 10^{-5}$, is almost the same with and without the time-delay interferometry combination.
This analysis enhances our understanding and underscores the crucial role of observations in advancing gravitational phenomena within String Theory.
\end{abstract}

\maketitle
\end{CJK*}

\section{Introduction}
In 2015, the Laser Interferometer Gravitational-Wave Observatory (LIGO) Scientific Collaboration and the Virgo Collaboration \cite{Abbott:2016blz, TheLIGOScientific:2016agk} made history by directly observing the first gravitational wave (GW) event, GW150914, originating from the merger of binary black holes (BHs).
This groundbreaking discovery opened a new window for exploring the properties of black holes and the nature of gravity in nonlinear and strong-field regimes \cite{Barsanti:2022vvl, Guo:2022euk, Maselli:2021men, Maselli:2020zgv, Barsanti:2022ana, Yunes:2011aa, Cardoso:2011xi, Zhang:2022rfr, AbhishekChowdhuri:2022ora, Torres:2020fye, Zhang:2022hbt, Zhang:2023vok, Liang:2022gdk, Bhattacharyya:2023kbh}.
Since then, dozens of GW events within the frequency range of tens to hundreds of Hertz have been confirmed \cite{LIGOScientific:2018mvr,LIGOScientific:2020ibl,LIGOScientific:2021usb,LIGOScientific:2021djp,LIGOScientific:2017vwq,LIGOScientific:2020aai,LIGOScientific:2021qlt}.
However, ground-based GW observatories are limited by seismic noise and gravity gradient noise, restricting their sensitivity to transient GWs in the frequency range of 10 to 1000 Hz, primarily emitted during the coalescence of stellar-mass compact binaries.
To overcome these limitations and explore a broader spectrum of GW sources, future space-based GW detectors, such as the Laser Interferometer Space Antenna (LISA) \cite{Danzmann:1997hm,LISA:2017pwj}, TianQin \cite{Luo:2015ght}, and Taiji \cite{Hu:2017mde}, will play a crucial role.
One of the most promising sources for future space-based GW detectors is extreme mass-ratio inspirals (EMRIs) \cite{Amaro-Seoane:2007osp, Babak:2017tow}. 
EMRIs consist of a stellar-mass compact object (secondary object) with a mass ranging from approximately 1 to 100 $M_{\odot}$ (e.g., black holes, neutron stars, white dwarfs) orbiting a SMBH with a mass between $10^5$ and $10^7$ $M_{\odot}$. 
These detectors will provide unprecedented insights into more GW sources, deviation from black hole and fundamental physics \cite{Gong:2021gvw,TianQin:2020hid,Ruan:2018tsw,LISA:2022yao,LISA:2022kgy,Ghosh:2024arw, Ghosh:2024arw, AbhishekChowdhuri:2023rfv, Rahman:2023sof, Rahman:2021eay, Rahman:2022fay, Rahman:2023sof, AbhishekChowdhuri:2023rfv, AbhishekChowdhuri:2023gvu, Kumar:2024utz}.

Unlike ground-based detectors, space-based GW detectors have the advantage of detecting long-duration GWs. However, the movement of spacecraft in space makes it difficult to maintain equal arm lengths, a challenge that complicates the cancellation of laser frequency noise. Because the arm lengths in space-based GW detectors are unequal, the laser frequency noise cannot simply cancel out.
To address this issue, a technique called time-delay interferometry (TDI) was developed \cite{Tinto:1994kg,Tinto:1999yr,Tinto:1999yr,Estabrook:2000ef,Tinto:2003vj,Vallisneri:2004bn}.
TDI involves carefully chosen time-shifted and linear combinations of signals to create virtual interferometric measurements with effectively equal arm lengths. The first-generation TDI combinations are capable of canceling out laser frequency noise in a static unequal-arm configuration, while the second-generation TDI extends this capability to time-dependent arm interferometry, further reducing frequency noise.
First-generation TDI \cite{Estabrook:2000ef,Tinto:2001ii, Tinto:2001ui,Hogan:2001jn,Armstrong:2001uh,Prince:2002hp,Tinto:2002de,Shaddock:2003dj,Tinto:2003uk,Nayak:2003na,Tinto:2004nz,Romano:2006rj,Zhang:2020khm} has been applied to space-based GW detectors like LISA, TianQin and Taiji, and discussions regarding the application of second-generation TDI configurations have also taken place \cite{Tinto:2003vj,Cornish:2003tz,Vallisneri:2004bn,Krolak:2004xp,Vallisneri:2005ji,Wang:2017aqq,Wang:2020fwa,RajeshNayak:2004jzp,Nayak:2005un}.

Although general relativity has accurately described current observations with high precision, both theoretical and experimental challenges suggest the need for an extension of this framework.
Key issues include its failure to account for the accelerating expansion of the Universe, its inability to incorporate dark matter, and its incompatibility with quantum field theory.
Observations of supermassive black holes (SMBHs) offer a valuable opportunity to test and constrain the effects of modified gravity theories. 
A promising approach is to explore deformed black holes by observing the GWs from EMRIs, where these deformations are parameterized and may hint at interpretations beyond general relativity.
For instance, in String Theory, such deformations are often associated with corrections scaled by $\alpha$, which is inversely related to the string tension \cite{Agurto-Sepulveda:2022vvf, Agurto-Sepulveda:2024iwu, Cano:2021rey}.
String Theory is inherently formulated in ten dimensions, necessitating a compactification process to link the theory to observable four-dimensional physics.
This compactification introduces higher-dimensional operators that augment the four-dimensional Einstein-Hilbert action.
These added operators perturbatively alter the dynamics and the structure of the solutions derived from the theory.

This paper aims to investigate how these perturbative modifications influence the EMRIs, providing insight into how String Theory's higher-dimensional corrections impact our understanding of these cosmic phenomena.
We consider the EMRI system that a stellar-mass BH inspires into a slowly rotating SMBH.
We obtain the evolution of the orbit by implementing the post-Newtonian GW energy flux and angular momentum flux and then constructing the waveforms.
The ability to constrain the String theory parameter $\alpha$ for detectors is calculated by the Fisher information matrix method.
We take LISA as a representative of the space-based detectors in our discussion,
and the analysis can be easily extended to other space-based detectors.
The paper is organized as follows. 
In Sec.~\ref{method}, we introduce the basic formalism of the slowly rotating black hole from String Theory and give the corrected orbital energy and angular momentum, as well as GW energy and angular momentum fluxes.
Then, we calculate the orbital evolution and waveform for EMRIs and compare them with the results from general relativity in Sec.~\ref{orbitandwaveform}.
In Sec.~\ref{TDI}, we provide a brief overview of TDI response and the Fisher information matrix (FIM) method.
The results of parameter estimation with the FIM method are shown in Sec.~\ref{result}. 
Last, we summarize in Sec.~\ref{conclusion}.

\section{The background metric and motion equation}\label{method}
The spacetime of the slowly rotating black hole coming from String Theory is \cite{Agurto-Sepulveda:2022vvf, Agurto-Sepulveda:2024iwu, Cano:2021rey}
\begin{equation}
\begin{split}
g_{tt} & =-1+\frac{2M}{r}+\frac{2Ma^{2}\cos^{2}\theta}{r^{3}}-\alpha f_{1}(r),\\
g_{t\phi} & =-\frac{2Ma\sin^{2}\theta}{r}+a\alpha h_{1}(r,\theta),
\end{split}
\end{equation}
\begin{equation}
\begin{split}
g_{rr} & =\frac{1}{1-\frac{2M}{r}+\alpha g_{1}(r)}-a^{2}\frac{(2M-r)\cos^{2}\theta+r}{(2M-r)^{2}r},\\
g_{\theta\theta} & =r^{2}+a^{2}\cos\theta,\\
g_{\phi\phi} & =(r^{2}+a^{2}+\frac{2Ma^{2}\sin^{2}\theta}{r})\sin^{2}\theta,      
\end{split}
\end{equation}
where
\begin{equation}
\begin{split}
f_{1}(r) & =\frac{8M^{2}}{r^{4}}+\frac{10M}{3r^{3}}+\frac{4}{r^{2}}-\frac{4}{Mr},\\
g_{1}(r) & =-\frac{40M^{2}}{3r^{4}}+\frac{2M}{r^{3}}+\frac{2}{r^{2}},\\
h_{1}(r,\theta) & =\left(\frac{8M^{2}}{r^{4}}+\frac{6M}{r^{3}}+\frac{6}{r^{2}}\right)\sin^{2}\theta\,.
\end{split}
\end{equation}
The geodesic motion around the slowly rotating black hole reads
\begin{equation}
\frac{d^{2}x^{\alpha}}{d\tau^{2}} + \Gamma^{\alpha}_{\beta} {}_{\gamma} \frac{dx^{\beta}}{d\tau}\frac{dx^{\gamma}}{d\tau}=0 ,
\end{equation}
where $\Gamma^{\alpha}_{\beta} {}_{\gamma}$ are the Christoffel symbols and $\tau$ is the affine parameter.
As we will consider equatorial orbits, that sets $\theta=\pi/2$.
The energy $E$ and the $z$ component of angular momentum $L^z$ are defined by \cite{Xin:2018urr}
\begin{equation}\label{ELzDef}
\begin{split}
-1=g_{\mu\nu}u^\mu u^\nu,\\
E=-u_t=-g_{tt}u^t-g_{t\phi}u^\phi,\\
L^z=u_\phi=g_{t\phi}u^t+g_{\phi\phi}u^\phi,
\end{split}    
\end{equation}
where $u^\mu=\frac{d x^\mu}{d\tau}$ is the $4$-velocity.
To facilitate calculations, we perform a variable transformation for bounded orbits
\begin{equation}
r(\chi)=\frac{M p}{1+e \cos(\chi)},
\end{equation}
where $p$ is the dimensionless semi-latus rectum and $e$ is the eccentricity.
The distances of the orbit's pericenter and apocenter are respectively given by
\begin{equation}
\begin{split}
r_{\rm apo}=\frac{Mp}{1-e},\\
r_{\rm peri}=\frac{Mp}{1+e}.
\end{split}
\end{equation}
Let's analyze the relationship between $(p,e)$ and $(E,L^z)$.
Since we have $dr/d\tau=0$ at $r_{\rm apo}$ and $r_{\rm peri}$, we can obtain the relationship between $(p,e)$ and $(E,L^z)$ from Eq.~\eqref{ELzDef}
\begin{equation}\label{EL}
\begin{split}
E&=E_{\rm GR}+\alpha~\delta E, \\
L^z&=L^z_{\rm GR}+\alpha~\delta L^z,
\end{split}
\end{equation}
where
\begin{equation}
\delta E=-\frac{\left(e^2-1\right) \left(68 e^4+e^2 (p (5 p-24)-136)-(p-2) (p (6 (p-6) p+29)+34)\right)}{6 p^{5/2} \sqrt{(p-2)^2-4 e^2} \left(-e^2+p-3\right)^{3/2}},
\end{equation}
\begin{equation}
\delta L^z=-\frac{\left(24 e^4+e^2 (17 p (p+4)+48)+p (3 p (1-2 (p-2) p)+28)-72\right)}{6 p^2 \left(-e^2+p-3\right)^{3/2}}.
\end{equation}
There are two frequencies $\Omega_r$ and $\Omega_\phi$ that correspond to the parameters $r(\chi)$ and $\phi$.
The coordinate time spent by the test body in moving from one pericenter to the next one is given by \cite{Fujita:2009bp}
\begin{equation}
\Lambda_t=\int_0^{\Lambda_t} dt=\int_0^{2\pi} \frac{dt}{d\chi}d\chi.     
\end{equation}
The corresponding change in the azimuthal angle $\phi$ reads
\begin{equation}
\Lambda_\phi=\int_0^{2\pi} \frac{d\phi}{d\chi}d\chi.
\end{equation}
The average angular velocity of $r(\chi)$ and $\phi$ over each period can be expressed as
\begin{equation}
\Omega_r=\frac{2\pi}{\Lambda_t},\qquad \Omega_\phi=\frac{\Lambda_\phi}{\Lambda_t},
\end{equation}
where
\begin{equation}
\Lambda_t=\Lambda_t^{\rm GR}+\alpha~\delta \Lambda_t,\qquad\Lambda_\phi=\Lambda_\phi^{\rm GR}+\alpha~\delta \Lambda_\phi,
\end{equation}
$\delta \Lambda_t$ and $\delta \Lambda_\phi$ are too lengthy to be fully written out here.
We use the quadrupole approximation to calculate the energy flux and angular momentum flux
\begin{eqnarray}
		\frac{d E}{d t}&=&\frac{1}{5}\left\langle\dddot{\mathcal{Q}}_{i j} \dddot{\mathcal{Q}}_{i j}\right\rangle,\\
		\frac{d L^{i}}{d t}&=&\frac{2}{5} \epsilon^{i k l}\left\langle\ddot{\mathcal{Q}}_{k a} \dddot{\mathcal{Q}}_{l a}\right\rangle,
\end{eqnarray}
where the quadrupole moment $\mathcal{Q}^{ij}$ can be described in terms of the mass moment $\mathcal{M}$ as follows
\begin{eqnarray}
	\mathcal{Q}^{i j}&=& \mathcal{M}^{i j}-\frac{1}{3} \delta^{i j} \mathcal{M}_{k k}, \label{QM}\\
	\mathcal{M}^{i j}&=&\int d^{3} x T^{00}(t, \mathbf{x}) x^{i} x^{j},
\end{eqnarray}
where $T^{ab}$ is the stress-energy tensor and $\mathbf{x}$ is the position vector.
In the weak field approximation, the leading-order energy flux and angular momentum flux are given as follows
\begin{equation}
\left\langle \frac{dE}{dt}\right\rangle=\dot{E}_{\rm GW}=\left\langle\frac{dE}{dt}\right\rangle_{\rm GR}+ \alpha\frac{ \mu ^2}{M^2} \frac{2\left(1-e^2\right)^{3/2} \left(37 e^4+292 e^2+96\right)}{5 p^5}.   
\end{equation}

\begin{equation}
\left\langle \frac{dL^z}{dt}\right\rangle=\dot{L}^z_{\rm GW}=\left\langle\frac{dL^z}{dt}\right\rangle_{\rm GR}+ \alpha\frac{ \mu ^2}{M^2}\frac{4 \left(1-e^2\right)^{3/2} \left(7 e^2+8\right)}{p^{7/2}}.   
\end{equation}

\section{Orbits and Waveforms}\label{orbitandwaveform}
Our method for calculating the bound orbital evolution due to radiation reaction is as follows. Given that the particle's motion is geodesic over a time scale comparable to the orbital period, we employ the adiabatic approximation, where the radiation reaction occurs over a much longer time scale. 
The orbital parameters change can be derived as follows
\begin{equation}
\label{balance}
    \left\langle \frac{d E}{dt} \right \rangle=\dot{E}_{\rm GW}=-\mu\dot{E},\qquad \left\langle \frac{d L^z}{dt} \right\rangle=\dot{L}^z_{\rm GW}=-\mu\dot{L}^z.
\end{equation}
Using Eq.~\eqref{balance}, we can infer the time-averaged change rates of the orbital parameters.
Since $E$ and $L^z$ are functions of $p$ and $e$, we have
\begin{equation}\label{balance2}
\begin{split}
   -\dot{E}_{\rm GW}&=\mu\frac{\partial E}{\partial p}\frac{dp}{dt}+\mu\frac{\partial E}{\partial e}\frac{de}{dt},\\
-\dot{L}^z_{\rm GW}&=\mu\frac{\partial L^z}{\partial p}\frac{dp}{dt}+\mu\frac{\partial L^z}{\partial e}\frac{de}{dt}.
\end{split}
\end{equation}
These equations can easily be inverted to get the orbital evolution $\{p(t),e(t)\}$,
\begin{equation}
\begin{split}
\frac{dp}{dt}&=\left[-\frac{\partial L^z}{\partial e}\dot{E}_{\rm GW}+\frac{\partial E}{\partial e}\dot{L}^z_{\rm GW}\right]\bigg/\left[\frac{\partial L^z}{\partial e}\frac{\partial E}{\partial p}-\frac{\partial E}{\partial e}\frac{\partial L^z}{\partial p}\right],\\
\frac{de}{dt}&=\left[+\frac{\partial L^z}{\partial p}\dot{E}_{\rm GW}-\frac{\partial E}{\partial p}\dot{L}^z_{\rm GW}\right]\bigg/\left[\frac{\partial L^z}{\partial e}\frac{\partial E}{\partial p}-\frac{\partial E}{\partial e}\frac{\partial L^z}{\partial p}\right].
\end{split}
\end{equation}
The orbit-averaged trajectory is determined over time in terms of $\{p(t), e(t), \Phi_{\phi}(t)$, $\Phi_{r}(t)\}$.
The phases $\Phi_{\phi,r}$ are obtained by integrating the orbit's fundamental frequencies over time
\begin{equation}
\Phi_{\phi,r}=\int_0^tdt'\,\Omega_{\phi,r}\left(p(t'),e(t')\right).
\end{equation}
The improved Augmented Analytic Kludge (AAK) waveforms remain valuable despite the development of faster relativistic waveforms.
We integrate the trajectory data with the waveform generator from the original AAK model, as detailed in \cite{Katz:2021yft}.
The amplitudes $h^{+,\times}$ in general AAK waveform are defined as 
\begin{equation}
\begin{split}
h^+(t)&=\sum_{n}A_n^+,\\
h^\times(t)&=\sum_{n}A_n^\times,\\
\end{split}
\end{equation}
where
\begin{equation}
\begin{split}
    A_n^+ =& - [1 + (\hat{L} \cdot \hat{N})^2 ] [a_n \cos(2 \gamma) - b_n \sin(2 \gamma)]+[1 - (\hat{L} \cdot \hat{N})^2 ] c_n\ , \\
     A_n^\times =& 2 (\hat{L} \cdot \hat{N}) [b_n \cos(2 \gamma) + a_n \sin(2 \gamma)]\ ,    
\end{split}
\end{equation}
where $\gamma=\Phi_\phi-\Phi_r$, $\hat{L}$ is the angular momentum of the secondary and $\hat{N}$ identifies the source location.
The coefficients $(a_n,b_n,c_n)$ are then given by
\begin{equation}
\begin{split}
a_n = & -n \mathcal{A}\big[J_{n-2}(n e) - 2 e J_{n-1}(n e)+(2/n)J_n(n e) \nonumber\\
 & + 2e J_{n+1}(ne)-J_{n+2}(ne)\big]\cos[n\Phi_{\phi}(t)]\ ,\\
 b_n =& -n \mathcal{A} (1-e^2)^{1/2}[J_{n-2}(ne)-2J_{n}(n e) \nonumber\\
 & +J_{n+2}(ne)]\sin[n\Phi_{\phi}(t)]\ ,\\
 c_n =& 2 \mathcal{A} J_n(ne) \cos[n\Phi_{\phi}(t)]\ ,    
\end{split}
\end{equation}
where $J_{n}$ is the Bessel function of the first kind,
$\mathcal{A}= (M\Omega_\phi )^{2/3} m_p/ d_L $, with $d_L$ being the source
luminosity distance.

\begin{figure*}[htp]
  \centering
  \includegraphics[width=0.95\textwidth]{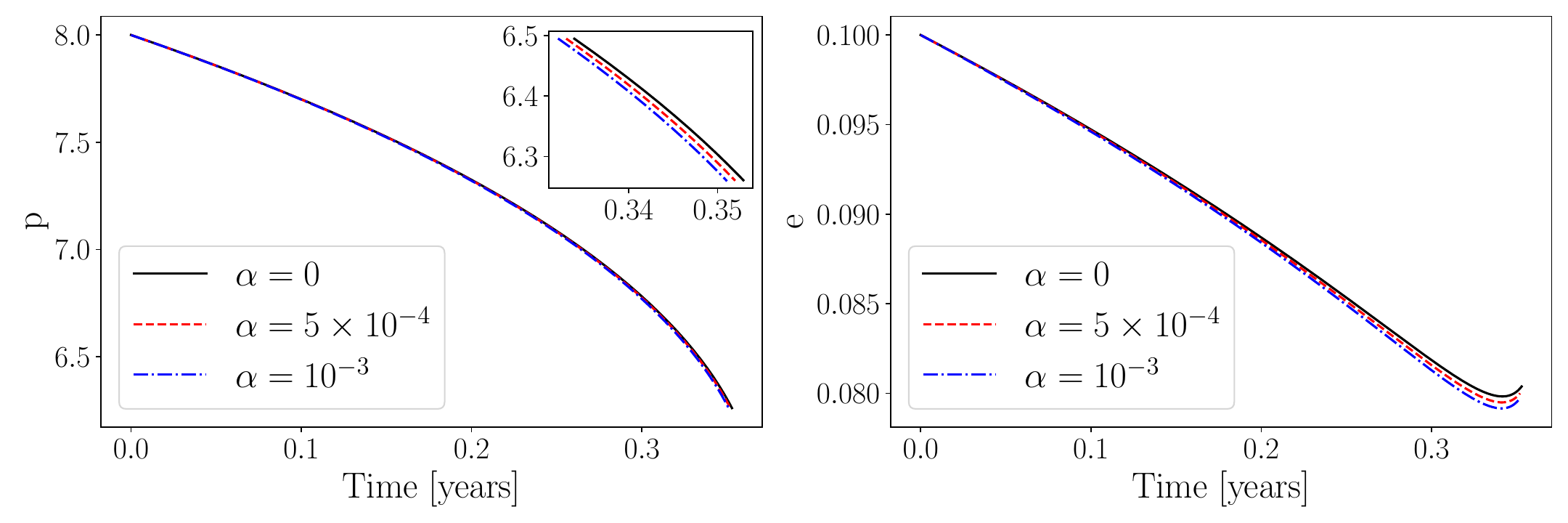}
  \caption{The orbital evolution for different values of $\alpha$. The initial conditions are chosen as $(a, p_0, e_0)$= $(0.1, 8.0, 0.1)$.
  The left panels show the evolution of the eccentricity $p$ with time, and the right panels show the evolution of $e$ with time.}
  \label{orbit}
\end{figure*}

\begin{figure*}[htp]
  \centering
\includegraphics[width=0.99\textwidth]{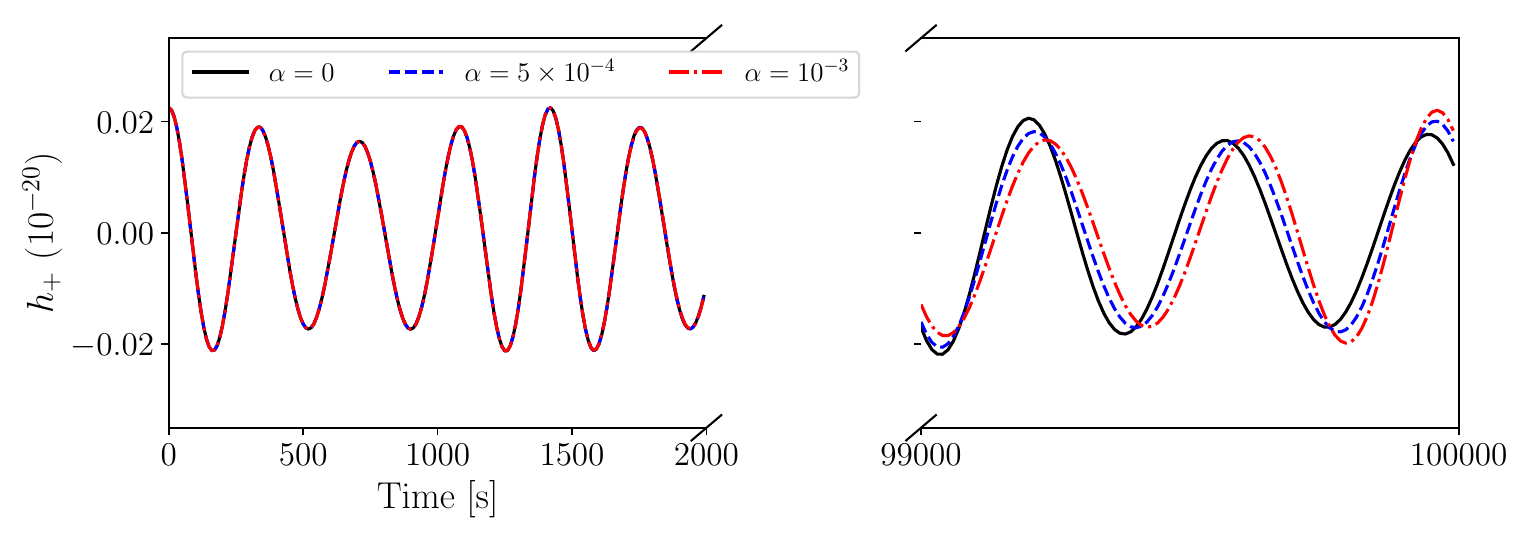}
  \caption{The waveforms with different values $\alpha$ for initial conditions $(a, p_0, e_0)$= $(0.1, 8.0, 0.1)$ after a long-time evolution.}
  \label{waveform}
\end{figure*}

We choose EMRI systems with $M=10^6~M_\odot$ for the central BH and $m_p=10~M_\odot$ for the small BH.
The orbital evolution with various $\alpha$ values is shown in Fig.~\ref{orbit}.
The initial conditions are chosen as $(a, p_0, e_0)$= $(0.1, 8.0, 0.1)$ before the orbital evolution starts.
We can see that the existing parameter $\alpha$ will accelerate the evolution of orbits because of the larger energy and angular momentum fluxes.
we assume that starting at $t=0$ the initial phase and positions of the small BH are the same for the situation with and without $\alpha$.
Figure~\ref{waveform} shows the GW waveforms with and without $\alpha$ after a long-time evolution from the initial position.
We observe distinct differences in GW between general relativity and String theory after a long-time evolution, which may help us constrain the String theory parameter $\alpha$.

\section{Time-delay interferometry and Fisher information matrix}\label{TDI}
Because the space-based detectors have three arms, they produce the same information as a pair of two-arm Michelson detectors.
The GW strains in the Solar system's barycenter (SSB) measured by the two Michelson detectors $\rm I$ and $\rm II$ are
\begin{equation}\label{signal}
h_{\rm{I}, \rm{II}}(t)=h_{+}^\text{SSB}(t) F^{+}_{\rm{I}, \rm{II}}(t)+h_{\times}^\text{SSB}(t) F^{\times}_{\rm{I}, \rm{II}}(t),
\end{equation}
where the interferometer pattern functions $F^{+,\times}_{\rm{I}, \rm{II}}(t)$ can be expressed in terms of four angles which specify the source orientation, $(\theta_s,\phi_s)$,
and the orbital angular direction $(\theta_1,\phi_1)$ \cite{Cutler:1997ta}.
The noise spectral density (PSD) for the two L-shaped Michelson detectors is provided by an accurate analytic fit, which only consists of the instrumental noise and ignores the confusion noise produced by unresolved galactic binaries \cite{Robson:2018ifk}
\begin{equation}
S_{\rm{I,II}}(f)=\frac{10}{3L^2}\left(1+\frac{6}{10}\frac{f^2}{f_*^2}\right)\left(P_{\rm oms}+2[1+\cos^2(f/f_*)]\frac{P_{\rm acc}}{(2\pi f )^4}\right),
\end{equation}
while
\begin{equation}
\begin{split}
        \sqrt{P_{\rm oms}} &= 15\times10^{-12} \sqrt{1+\left(\frac{2\times10^{-3}}f\right)^4} , \\
        \sqrt{P_{\rm acc}}&=3\times10^{-15} \sqrt{1+\left(\frac{0.4\times10^{-3}}f\right)^2}\sqrt{1+\left(\frac f{8\times10^{-3}}\right)^4}.      
\end{split}  
\end{equation}
where $L=2.5~\rm{Gm}$ and $f_*=19.09~\rm{mHz}$ are the arm distance and characteristic frequency of LISA.
In practice, maintaining exact equality of the arm lengths is challenging due to the motion of space-based spacecraft. 
As a result, laser frequency noise cannot be effectively removed when the two beams are directly recombined at the photo-detector.
By appropriately selecting time-shifted and linear combinations of these readouts, it is possible to mitigate these noise sources.
This method is known as time-delay interferometry (TDI).
The deformation $H_{12}(t)$ on the link unit vector $\hat{n}_{12}$ (from the emitter spacecraft $2$ to the receiver spacecraft $1$) can be calculated by projecting the GW onto the detector arm \cite{Baghi:2023qnq}
\begin{equation}
H_{12}(t) = h_+^\text{SSB}(t) \times \xi_+(\hat{u}, \hat{v}, \hat{n}_{12}) + h_\times^\text{SSB}(t) \times \xi_\times(\hat{u}, \hat{v}, \hat{n}_{12}),
\label{eq:projected-strain-ssb}
\end{equation}
and 
\begin{equation}
\begin{split}
    \xi_+(\hat{u}, \hat{v}, \hat{n}_{12}) &= (\hat{u} \cdot \hat{n}_{12})^2 - (\hat{v} \cdot \hat{n}_{12})^2, 
    \\
    \xi_\times(\hat{u}, \hat{v}, \hat{n}_{12}) &= 2 (\hat{u} \cdot \hat{n}_{12}) (\hat{v} \cdot \hat{n}_{12}),
\end{split}   
\end{equation}
where $\hat{u}$ and $\hat{v}$ are polarization vectors.
During the light travel time, the wave propagation time to first order is $t(\lambda) \approx t_2 + \lambda/c$ and the wave position to first order is $\hat{x}(\lambda) = \hat{x}_2(t_2) + \lambda \hat{n}_{12}(t_2)$, where $\hat{x}_2(t_2)$ denotes the position of the emitter spacecraft at the emission time and $\hat{x}_1(t_1) = \hat{x}_2(t_2) + L_{12} \hat{n}_{12}(t_2)$ denotes the position of the receiver spacecraft at the receiver time.
Under the approximation that the spacecraft moves slowly relative to the GW propagation timescale
\begin{equation}
\begin{split}
t_1 &\approx t_2 + L_{12} / c ,\\
\hat{x}_2(t_2) &\approx \hat{x}_2(t_1),\\
\hat{n}_{12}(t_1) &\approx \hat{n}_{12}(t_2).
\end{split}
\end{equation}
Finally, the relative frequency shift $y_{12}$, experienced by light as it travels from the emitter spacecraft $2$ to the receiver spacecraft $1$, is given
\begin{equation}
y_{12}(t_1) \approx \frac{1}{2  (1 - \hat{k} \cdot \hat{n}_{12}(t_1))} \Big[ H_{12} \Big(t_1 - \frac{L_{12}(t_1)}{c} - \frac{\hat{k} \cdot \hat{x}_2(t_1)}{c} \Big) - H_{12}  (t_1 - \frac{\hat{k} \cdot \hat{x}_1(t_1)}{c}) \Big],
\label{eq:instrument-response-to-gw}
\end{equation}
where $\hat{k}$ is the GW propagation vector, $L_{12}$ is the arm length between the emitter spacecraft $2$ and the receiver spacecraft $1$.
The other relative frequency shift, $y_{ij}$, encountered by light as it travels along link $ij$, can be combined in different ways to derive the TDI observables \cite{Katz:2022yqe}.
The first-generation Michelson TDI combinations, $X, Y, Z$ are given by~\cite{Tinto:2003vj},
\begin{equation*}
    \begin{split}
        X &=
        y_{13} + \delay{13} y_{31} + \delay{131} y_{12} + \delay{1312} y_{21} - [y_{12} + \delay{12} y_{21} + \delay{121} y_{13} + \delay{1213} y_{31}] ,\\ 
        Y &=
        y_{21} + \delay{21} y_{12} + \delay{212} y_{23} + \delay{2123} y_{32} - [y_{23} + \delay{23} y_{32} + \delay{232} y_{21} + \delay{2321} y_{12}] ,\\ 
        Z &=
        y_{32} + \delay{32} y_{23} + \delay{323} y_{31} + \delay{3231} y_{13} - [y_{31} + \delay{31} y_{13} + \delay{313} y_{32} + \delay{3132} y_{23}].
\end{split}
\end{equation*}
Delay operators are defined as follows
\begin{equation}
    \delay{i_1, i_2, \dots, i_n} x(t) = x\left(t - \sum_{k=1}^{n-1}{L_{i_k i_{k+1}}/c}\right).
\end{equation}
These combinations, $X, Y, Z$, have correlated noise properties \cite{Marsat:2020rtl}.
An uncorrelated set of TDI variables, $A, E, T$, can be obtained from linear combinations of $X,Y,Z$ given by
\begin{align}
    A =& \frac{1}{\sqrt{2}}\left(Z-X\right),  \\
    E =& \frac{1}{\sqrt{6}}\left(X-2Y+Z\right) ,\\
    T =&\frac{1}{\sqrt{3}}\left(X+Y+Z\right).
\end{align}
The noise PSD for $A, E, T$ takes the form \citep{Marsat:2020rtl}
\begin{equation}
    \begin{aligned}
        S_{A, E} &= 8\sin^2 (2 \pi f L)\left(
        {[2 + \cos (2 \pi f L)] S_{\rm oms} +
        [6 + 2\cos (4\pi f L) + 4\cos (2\pi f L)] S_{\rm acc}}\right), \\
        S_{T} &= 32 \sin^2 (2 \pi f L) \sin^2 (\pi f L)
        [S_{\rm oms} + 4 \sin^2 (\pi f L) S_{\rm acc}],
    \end{aligned}
\end{equation}
while
\begin{equation}
    \begin{aligned}
        \sqrt{S_{\rm oms}} &= 15\times10^{-12} \frac{2\pi f}c \sqrt{1+\left(\frac{2\times10^{-3}}f\right)^4} , \\
        \sqrt{S_{\rm acc}}&= \frac{3\times10^{-15}}{2\pi f c} \sqrt{1+\left(\frac{0.4\times10^{-3}}f\right)^2}\sqrt{1+\left(\frac f{8\times10^{-3}}\right)^4}.
    \end{aligned}
\end{equation}
The inner products are summed over the three channels,
\begin{equation}
\left\langle a|b \right\rangle = \sum_{i = A,E,T}\left\langle a^i|b^i \right\rangle.
\end{equation}
In the time domain, the GW signal is mainly determined by parameters
\begin{equation}
{\bm \xi}=(\ln M, \ln m_p, a, p_0, e_0, \alpha, \theta_s, \phi_s, \theta_1, \phi_1, d_L),
\end{equation}
where $p_0$ and $e_0$ are the radius and eccentricity of the smaller compact object at $t=0$.
Bayesian inference method is based on Bayes rule
\begin{equation}\label{EqBayes}
    p({\bm \xi}|d) = \frac{p(d|{\bm \xi})p({\bm \xi})}{p(d)},
\end{equation}
where $p({\bm \xi}|d)$ is the posterior distribution of the parameters ${\bm \xi}$, $p(d|{\bm \xi})$ is the likelihood,
\begin{equation}\label{EqLikelihood}
    p(d|{\bm \xi})=\exp\left[-\frac{1}{2}(h({\bm \xi})-d|h({\bm \xi})-d)\right],
\end{equation}
$d=h({\bm \xi_0})+n$ is the observed data for the true parameters ${\bm \xi_0}$,
$n$ is the noise generated by the noise power spectra,
$p({\bm \xi})$ is the prior distribution of the parameters ${\bm \xi}$,
and $p(d)$ is the evidence which is treated as a normalization constant.
In the large SNR limit,
the covariances of source parameters $\xi$  are given by the inverse of the Fisher information matrix
\begin{equation}
\Gamma_{i j}=\left\langle\left.\frac{\partial h}{\partial \xi_{i}}\right| \frac{\partial h}{\partial \xi_{j}}\right\rangle_{\xi=\hat{\xi}}.
\end{equation}
The statistical error on $\xi$ and the correlation coefficients between the parameters are provided by the diagonal and non-diagonal parts of ${\bf \Sigma}={\bf \Gamma}^{-1}$, i.e.
\begin{equation}
\sigma_{i}=\Sigma_{i i}^{1 / 2} \quad, \quad c_{\xi_{i} \xi_{j}}=\Sigma_{i j} /\left(\sigma_{\xi_{i}} \sigma_{\xi_{j}}\right).
\end{equation}
The total SNR can be written as 
\begin{equation}
\rho=\sqrt{\rho_A^2+\rho_E^2+\rho_T^2}=\sqrt{\left\langle h_A|h_A \right\rangle+\left\langle h_E|h_E \right\rangle+\left\langle h_T|h_T \right\rangle},
\end{equation}
where $h_{A,E,T}$ denote the TDI signals detected by the detectors.
The total covariance matrix of the source parameters is obtained by inverting the sum of the Fisher matrices $\sigma_{\xi_i}^2=(\Gamma_A+\Gamma_E+\Gamma_T)^{-1}_{ii}$.
Because the signal of $T$ channel is much less than the channels $A, E$, we will consider only $A, E$ channels \cite{Marsat:2020rtl}.

\section{Result}\label{result}
\begin{figure}
    \centering
    \includegraphics[width=0.9\columnwidth]{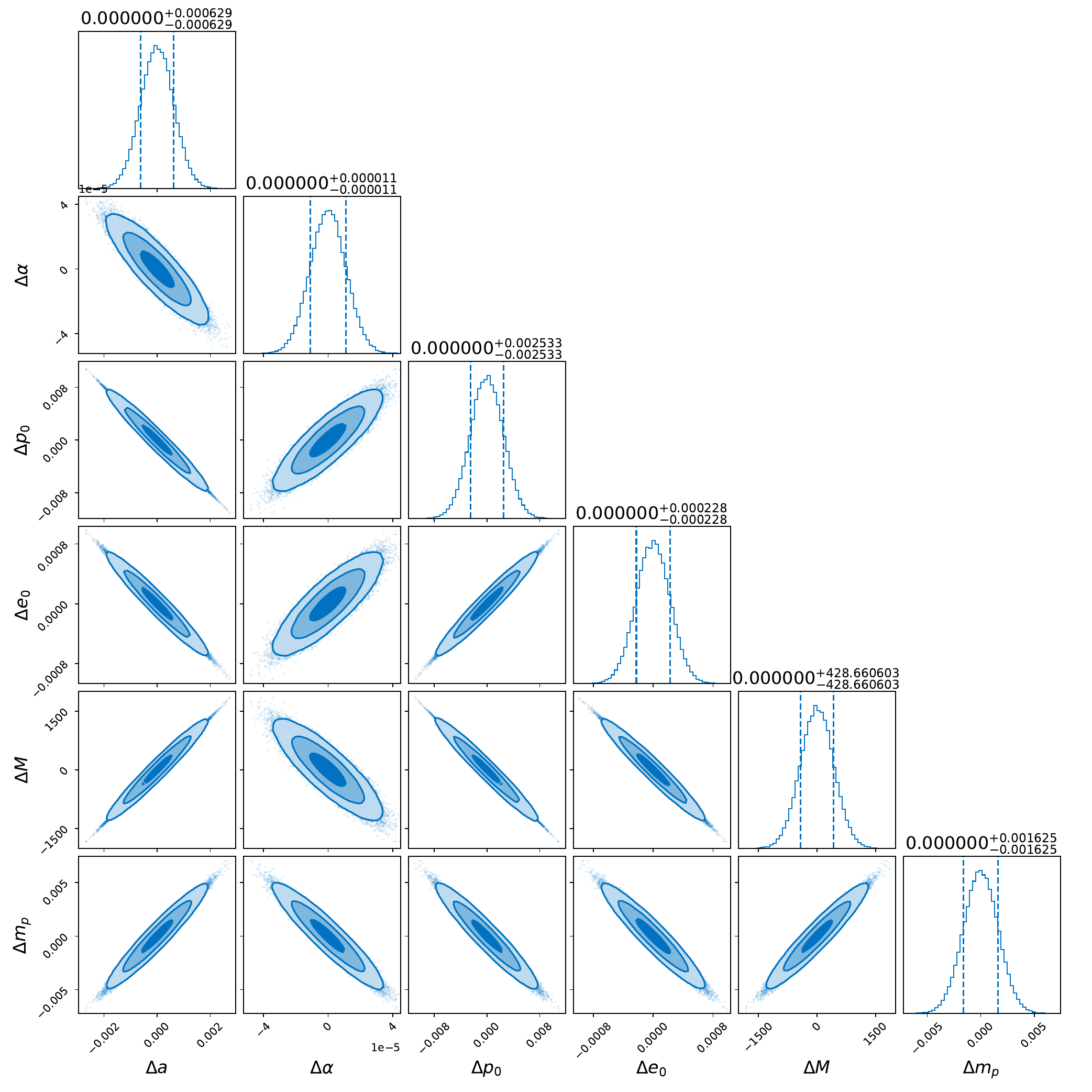}
    \caption{Corner plot for the probability distribution of intrinsic parameters for two Michelson space-based detectors. Diagonal boxes refer to marginalized distributions. Vertical lines show the 1-$\sigma$ interval for each waveform parameter. The contours correspond to the $68\%$, $95\%$, and $99\%$ probability confidence intervals.}
    \label{cornerMich}
\end{figure}

\begin{figure}
    \centering
    \includegraphics[width=0.9\columnwidth]{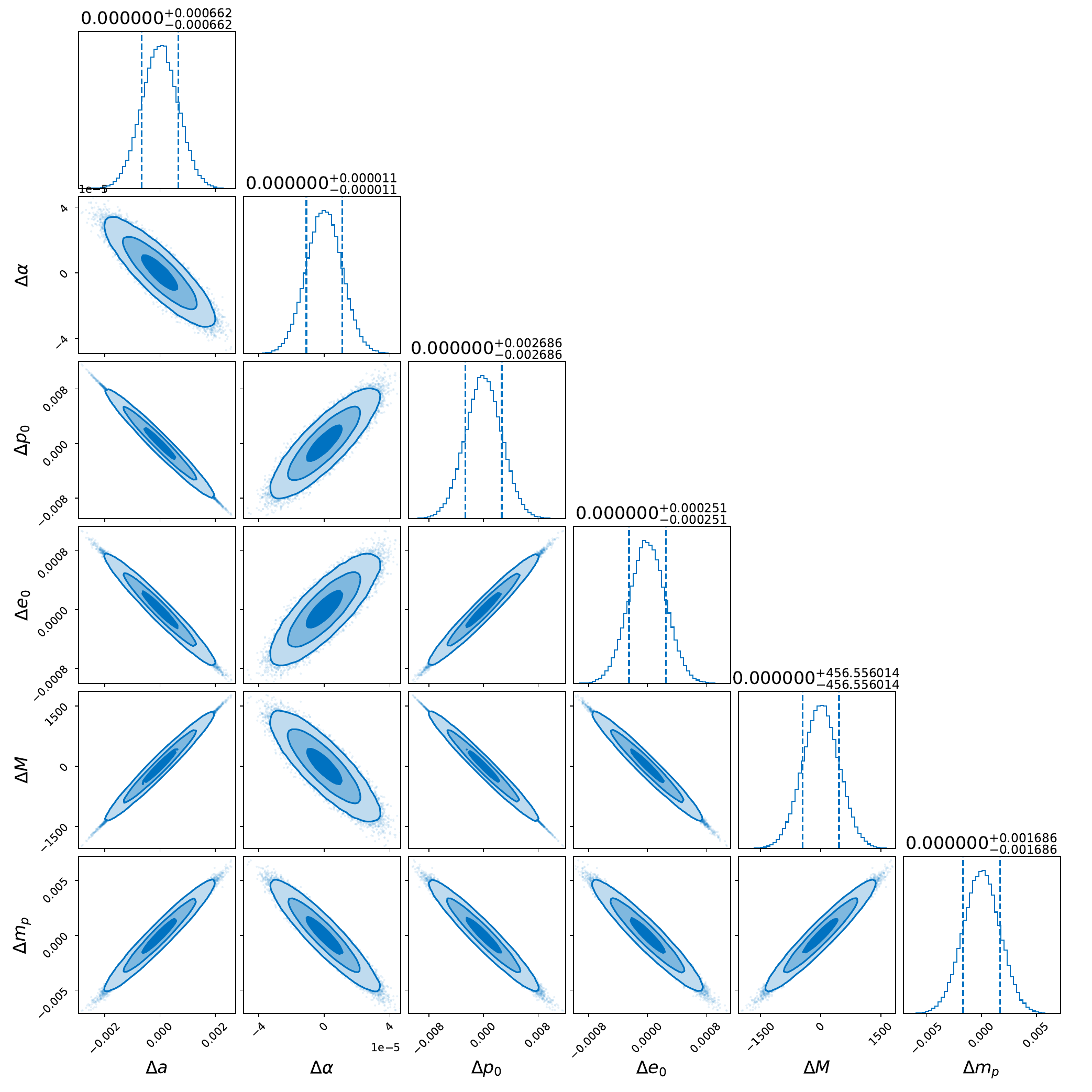}
    \caption{Corner plot for the probability distribution of intrinsic parameters for TDI $A,E,T$. Diagonal boxes refer to marginalized distributions. Vertical lines show the 1-$\sigma$ interval for each waveform parameter. The contours correspond to the $68\%$, $95\%$, and $99\%$ probability confidence intervals.}
    \label{cornerTDI}
\end{figure}

The source angles are fixed $\theta_s=\pi/3,~\phi_s=\pi/2$, $\theta_1=\pi/4,~\phi_1=\pi/4$, and the initial orbital separation $p_0$ and eccentricity $e_0$ is adjusted to experience one-year adiabatic evolution before the final plunge.
We perform the FIM estimation for the EMRI system with $m_p=10~M_{\odot}$, $M=10^6~M_{\odot}$, $a=0.1$, $\alpha=0$, $p_0=9.24419$, $e_0=0.1$, and the luminosity distance $d_L=1~\rm{Gpc}$.
Figure \ref{cornerMich} shows the probability distribution obtained by the Fisher matrix information for the binary masses, the spin of the primary, and the parameter $\alpha$, for EMRIs observed one year before the plunge and for two Michelson space-based detectors.
This analysis shows that the measurement of the parameter $\alpha$ can be detected with the error $\Delta\alpha=1.1\times 10^{-5}$ with SNR$=96$.
Figure \ref{cornerTDI} shows the probability distribution obtained by the Fisher matrix information for the binary masses, the spin of the primary, and the parameter $\alpha$, for EMRIs observed one year before the plunge and for TDI $A,E,T$.
This analysis shows that the measurement of the parameter $\alpha$ can be detected with the error $\Delta\alpha=1.1\times 10^{-5}$ with SNR$=104$.

\section{Conclusion}\label{conclusion}
For space-based GW detectors, the technique of TDI is needed to cancel out laser frequency noises due to the difficulty of maintaining the exact equality of the arm length in space. 
We give the deviation effect from string theory parameter $\alpha$ on gravitational wave fluxes, orbital evolution, and phase dynamics with leading-order post-Newtonian corrections.
The larger parameter $\alpha$ accelerates the evolution of $p$ and $e$ due to the larger energy and angular momentum carried away.
The Fisher information matrix method is applied to assess the possibility of detecting extreme mass ratio inspirals parameters as well as string theory parameter $\alpha$.
With one-year observations of LISA, we can constrain the parameter $\alpha$ less than about $10^{-5}$ with $\rm SNR$ around $100$.
The constraint on modified black holes with extreme mass ratio inspirals is almost the same with and without the TDI combination.
This analysis provides an understanding while highlighting the essential role of observations in advancing gravitational phenomena in the String Theory.

\begin{acknowledgments}
This work is supported in part by the National Key Research and Development Program of China under Grant No. 2020YFC2201504, and the China Postdoctoral Science Foundation under Grant No. 2023M742297.
\end{acknowledgments}


%

\end{document}